\documentclass[12pt]{iopart}
\usepackage{etoolbox}       % the following 4 lines are added by yxh
\csundef{equation*}
\csundef{endequation*}
\usepackage{amsmath}
\usepackage{graphicx}

\usepackage{color}
\usepackage{ulem}
\usepackage{enumerate}

\begin{document}

\title[]{Control of fast electron propagation in foam target by high-Z doping}

\author{H Xu$^{1,*}$, X H Yang$^{2,3,\dag}$, J Liu$^{1}$, and M Borghesi$^{4}$}

\address{$^1$School of Computer Science, National University of Defense Technology, Changsha 410073, China}
\address{$^2$Department of Physics, National University of Defense Technology, Changsha 410073, China}
\address{$^3$Department of Mechanical Engineering, University of Rochester, Rochester, New York 14627, USA}
\address{$^4$School of Mathematics and Physics, Queen's University of Belfast, Belfast BT7 1NN, United Kingdom}

\ead{xuhan$_{-}$email@aliyun.com}
\ead{xhyang@nudt.edu.cn}

\begin{abstract}
The influence of high-Z dopant (Bromine) in low-Z foam (polystyrene) target on laser-driven fast electron propagation is studied by the 3D hybrid particle-in-cell (PIC)/fluid code HEETS.
It is found that the fast electrons are better confined in doped targets due to the increasing resistivity of the target, which induces a stronger resistive magnetic field which acts to collimate the fast electron propagation.
The energy deposition of fast electrons into the background target is increased slightly in the doped target, which is beneficial for applications requiring long distance propagation of fast electrons, such as fast ignition.
\end{abstract}

\pacs{52.25.Fi, 52.25.Jm, 52.65.Ww} \submitto{\PPCF}

\maketitle

\section{Introduction}\label{sec:1}
Laser-driven fast electron propagation in over-dense plasmas has attracted great recent interests due to its applications in the fast ignition scheme of inertial confinement fusion \cite{Tabak94}, laser-driven particle acceleration \cite{Macchi13,Yang10} as well as ultrashort bright x-ray sources \cite{Jin11}. However, the fast electrons usually have a large divergence due to the scattering of laser transverse ponderomotive force \cite{Debayle10,Ovchinnikov13} and magnetic fields generated by the Weibel instability \cite{Adam06,Yang16}. It is important to control the fast electron propagation in dense plasmas in order to facilitate the aforementioned applications.

The generation of magnetic fields, as the fast electrons propagate in solid plasmas, is described by \cite{Davies97} $\partial_t\vec{B}=\eta\nabla\times\vec{J}_f+\nabla\eta\times\vec{J}_f$,
where $\eta$ is the target resistivity and $\vec{J}_f$ is the fast electron current density. The first term on the right hand side of the equation acts to focus the fast electrons to the region of higher current density, while the second term will push the fast electrons towards regions of higher resistivity, inducing beam hollowing provided that the resistivity decreases with temperature.
Thus, a possible route to control the fast electrons is by applying a resistivity gradient in the background plasma.
This can be achieved, for example, by engineering targets that have higher resistivity around the fast electron propagation axis, so that the fast electrons can be pushed towards the axis.
A large variety of targets specially designed to generate such collimated magnetic fields have been proposed recently, such as targets having higher resistivity core and lower resistivity cladding \cite{Robinson07} or Switchyard targets \cite{Robinson12}, which can generate strong magnetic fields around the interface of the materials and show an effective collimation of the fast electrons.
It has also been suggested that control of fast electron propagation in metal targets could be possible by tuning the target ionizations dynamics, which affects the resistive magnetic field growth and then provides feedback on the fast electron propagation \cite{Sentoku11}, and more collimated fast electron beams are obtained in high-Z targets.
Collimation of fast electrons in compressed targets is also observed due to the fact that the self-generated magnetic field can have longer time to grow in such targets due to the higher specific heat capacity compared to solid targets \cite{Perez11,Yang12}.
It is also noted that the target resistivity at low temperature as well as the target's lattice structure can have a significant effect on the fast electron propagation, which can lead to fast electron filamentation or annular propagation depending on the resistivity profiles \cite{McKenna11,MacLellan13}. Therefore, describing the target characteristics (ionization, resistivity, specific heat capacity, et al.,) accurately is essential to correct modeling of fast electron propagation in dense plasmas.

In this paper, fast electron propagation in plastic foam target and foams doped with different percentages of Br in order to modulate the target resistivity are studied by hybrid PIC/fluid simulations. It is found that the fast electrons can be better collimated in doped targets due to the higher resistivity and stronger resistive magnetic field. The filamentation of fast electrons is also mitigated in the doped targets. The paper is organized as follow: Section II describes the physical model employed in the hybrid PIC/fluid code HEETS \cite{Xu17,Yang18}. Section III shows the transport parameter models used in the work. Section IV introduces the simulation model of this work. Section V presents the simulation results of fast electron propagation in foam targets and doped targets. Finally, a conclusion is presented.

\section{Hybrid model of HEETS}\label{sec:2}
To model relativistic electron beams propagating in cold dense plasmas, both effects of electromagnetic fields and particle collisions should be considered appropriately. It is helpful to describe the fast electrons and background plasma separately in order to capture the key physics and make it possible to be modeled with current computing resources.
We have newly developed a 3D parallel PIC/fluid hybrid code named HEETS.
It treats the fast electrons by a collisional and fully electromagnetic method (i.e., explicit PIC), while background electron-ion plasmas are described by a reduced two-fluids model, which is similar to that of Davies \cite{Davies09}, Zephyros \cite{Robinson15}, and Zuma \cite{ZUMA}.

The fast electrons are advanced according to the equations
\begin{equation}\label{pic}
d\vec{x}/dt=\vec{v}, \;\;
d\vec{p}/dt=-e(\vec{E}+\vec{v}\times\vec{B}) + {\delta\vec{p}},
\end{equation}
where $\vec{x}$ is the electron position and $\vec{p}=\gamma m_e\vec{v}$ is the relativistic momentum. The term $\delta\vec{p}$ represents the fast electron energy loss and angular scattering due to the collision with the background particles. The other symbols have their usual meanings. Note that Eq. (\ref{pic}) is valid only when the speed of fast electrons is much greater than the averaged speed of the background particles, and its density is much less than that of the latter.

It is convenient to separate $\delta\vec{p}$ into drag and scattering terms
\begin{equation}\label{sde}
dp=-\frac{DL_d}{v^2}dt, \;\;
d\theta=\sqrt{\frac{2ZDL_s}{\gamma^2m_ev^3}}dW.
\end{equation}
where $p$ is the amplitude of fast electron momentum, $\theta$ is the scattering angle, $D=n_{eb}^{tot}e^4/4\pi\epsilon_0^2m_e$, $n_{eb}^{tot}$ is the total background electron density (including free and bound electrons), $Z$ is the atomic number. The sample trajectories can be obtained by expressing $dW=G(t)dt^{1/2}$, where $W$ is the distribution of the Weiner process, $G(t)$ is a random variable of Gaussian distribution with mean zero and variance one.
It is noted that the drag number $L_d$ and scattering number $L_s$ differ slightly for different semi-empirical models \cite{Davies97,Robinson15,ZUMA,Atzeni09},
which are weakly dependent on the material and fast electron energy and have typically values of 5-20.
It is seen that there is no obvious difference for using different drag and scattering models for our simulation here.
Equations (\ref{pic}) and (\ref{sde}) can be solved numerically using a `split step operator' method \cite{McLachlan02}.
Firstly, the momentum of fast electron $\vec{p}^{old}$ is updated to $\vec{p'}$ using the Boris-rotation algorithm \cite{Birdsall85} without collision correction $\delta\vec{p}$.
Secondly, the temporary momentum $\vec{p'}$ is transformed to a local frame by rotating it with the scattering angle, in which the $z$ axis of the frame aligns to the direction of $\vec{p'}$, and then updated to $\vec{p'}^{new}$ using Eq. (\ref{sde}).
Finally, the momentum $\vec{p'}^{new}$ is transformed back to the laboratory frame to obtain the final electron momentum $\vec{p}^{new}$.
After the fast electron momentum and position are updated,
the current density $\vec{J_f}$ is deposited onto the spatial grid using a  charge-conserved current deposition algorithm \cite{Umeda}.
A 5th-order spline spatial interpolation algorithm \cite{Abe} is employed to deposit the current, which allows larger spatial grids and time steps to be applied.

The background current density $\vec{J_b}$ is obtained from Ampere's law with displacement current ignored
\begin{equation}\label{ampere}
\vec{J_b}= -\vec{J_f} + \mu_0^{-1}\bigtriangledown\times\vec{B},
\end{equation}
The magnetic field $\vec{B}$ is updated by Faraday's law
\begin{equation}\label{faraday}
\partial\vec{B}/\partial{}t= -\bigtriangledown\times\vec{E},
\end{equation}
The electric field $\vec{E}$ is given by Ohm's law without advection term $\vec{}v_{eb}\times\vec{B}$ and thermal pressure,
\begin{equation}\label{ohm}
\vec{E}= \eta\vec{J_b}.
\end{equation}
where $\vec{v}_{eb}$ are the velocity of fluid element of electrons, and the resistivity $\eta$ depends on the material, electron density $n_{e}$ and temperature $T_{e}$, and ion temperature $T_{i}$.

The background electron temperature $T_{e}$ (in units of eV) and the ion temperature $T_{i}$ (in units of eV) are determined by
\begin{subequations}\label{temp}
\begin{equation}\label{tempa}
\partial(C_eT_{e})/\partial{t}= \vec{E}\cdot\vec{J_b} + \bigtriangledown\cdot(\kappa\bigtriangledown T_{e}) + Q_f + Q_{ie},
\end{equation}
\begin{equation}\label{tempb}
\partial(C_iT_{i})/\partial{t}= Q_{ei}.
\end{equation}
\end{subequations}
The electron specific heat capacity $C_e$ is calculated by $C_e= (C_{e1}^{-2}+C_{e2}^{-2})^{-1/2}$, as reported in Ref. \cite{Fisher01,Antici13}, where $C_{e1}=\frac{1}{2}\pi^2n_{e}T_{e}/T_F$ is the electron capacity for a degenerate plasma, $C_{e2}=\frac{3}{2}n_{e}$ is the electron heat capacity for a Maxwellian plasma with $n_{e}=Zn_{i}$,
where $T_F$ is the Fermi temperature, $n_{e}$ is the free electron density, $n_{i}$ is the ion density, $Z$ is the ionization degree.
Though the above specific capacities are quite simple, it has been proved that this approach can reproduce successfully the SESAME data \cite{Eidmann00}.
For the case $T_{e}>>T_F$, the specific heat capacity $C_e\approx\frac{3}{2}n_{e}$ is widely used in hybrid codes \cite{Davies09,Robinson15,ZUMA}.
The ionization degree $Z$, as well as resistivity $\eta$, having an important impact on hybrid simulations, will be discussed later in detail.

The 1st term $\vec{E}\cdot\vec{J_b}$ on the RHS of Eq. (\ref{tempa}) is the Ohmic heating, which plays a dominant role in fast electron propagation in solid targets as considered here.
The 2nd term $\bigtriangledown\cdot(\kappa\bigtriangledown T_{e})$ is thermal conduction, where $\kappa$ is the thermal conductivity. This term becomes important only when electron temperature gets high and usually requires a relatively long time evolution.
The Spitzer thermal conductivity model \cite{Salzmann98}  $\kappa\equiv\kappa_S=G(Z)\cdot
\frac{10.16k_B(4\pi\epsilon_0)^2T_{e}^{5/2}}{Ze^4m_e^{1/2} \ln\Lambda}$ is reasonable in most cases, where $G(Z)=\frac{0.47Z}{Z+4}$ is the electron-electron collision correcting factor, $\ln\Lambda$ is the Coulomb logarithm with typical value of $5\sim10$, $k_B$ is the Boltzmann constant.
The 3rd term $Q_{f}=\frac{3}{2}T_fn_f/\tau_{fb}$ is the fast-electron frictional energy loss \cite{Michael95}, where $T_f=\frac{1}{3}m_e(\langle{}v_f\rangle{}^2-\langle{}v_f^2\rangle{})$ is the temperature of fast electrons, $n_f$ is the number density of fast electrons, $\tau_{fb}=\frac{3\sqrt{3}}{8\pi}\frac{(4\pi\epsilon_0)^2m_e^{1/2}T_f^{3/2}}{n_{e}e^4ln\Lambda}$ is the relaxation time due to collisions of fast electron with background electron.

$Q_{ie}$ and $Q_{ei}$ are the ion-electron and electron-ion energy transfer terms via Coulomb collisions, respectively. The energy transfer by Coulomb collision for species `a' to species `b' is described by Landau formula \cite{Spitzer62} $Q_{ab}=3m_b\nu_{ba}n_a(T_b-T_a)$,
where $\nu_{ba}=\frac{4}{3}(\frac{2\pi}{m_b})^{1/2}(\frac{q_aq_b}{4\pi\epsilon_0})^2\frac{n_aL_{ba}}{(T_b+T_am_b/m_a)^{3/2}}$
is the collisional rate.
The Coulomb logarithm is given by $L_{ba}=\ln(r_d/b_0)$,
where $r_d=r_{da}r_{db}/\sqrt{(r_{da}^2+r_{db}^2)}$,
$r_{dj}= v_{Tj}/\omega_{pj}$,
$v_{Tj}$ and $\omega_{pj}$ are the thermal velocity and plasma frequency of species `j',
$b_0=q_aq_b/[3(4\pi\epsilon_0)(T_a+T_b)]$ is the impact parameter of Coulomb collision.

Equations (\ref{ampere})-(\ref{temp}) can be solved using an iterative method. The step from time $n$ to $n+1$ consists of:
\begin{enumerate}[1)]
\item $\vec{B}$ is updated to $n+1$ time step using Eq. (\ref{faraday}).
\item $\vec{J_b}$ is updated to $n+1$ time step using Eq. (\ref{ampere}).
\item $\vec{E}$ is updated to $n+1$ time step using Eq. (\ref{ohm}).
\item $T_{e}, T_{i}$ are updated to $n+1$ time step using Eq. (\ref{temp}).
\item The transport parameters $C_e, C_i, \eta, \kappa$ are updated to $n+1$ time step using $T_{e}^{n+1}, T_{i}^{n+1}$.
\item Goto step 1) until the variation of all fields $\vec{B}, \vec{E}, T_{e}, T_{i}$ is within a prescribed margin.
\end{enumerate}

It is shown that the relative variation of the fields will reduce to less than $10^{-5}$ after 5-10 iterations. In addition, the predictor-corrector method is also applied to numerically solve Eqs. (\ref{ampere})-(\ref{temp}) \cite{Xu17}.

\section{Transport parameter models}\label{sec:3}
The accuracy of transport parameters, especially for the ionization degree $Z$ and resistivity $\eta$, plays a key role in simulations of the fast electron propagation in cold dense matter. Generally, these transport parameters are function of the temperature ($T_{e}$ and $T_{i}$) and density $\rho$ of the background material.
HEETS reads the parameters from several pre-prepared tables (one for each material) during the simulation.
Each table includes the value of averaged ionization degree $\bar{Z}$ (for compounds), resistivity $\eta$, and thermal conductivity $\kappa$ on discrete sample points of density $\rho$ and electron temperature $T_{e}$. These tables can be generated by pre-processing with HEETS or by other programs.

The transport parameters in warm dense matter (WDM) have attracted significant recent attention. Some numerical methods, such as density functional theory-based quantum molecular dynamics (DFT-QMD) \cite{Marques12,Wolfram01}, or averaged atoms model \cite{Liberman79,Wilson06,Blenski95}, can provide more accurate transport parameters. However, such models are usually complex and time consuming in realistic calculations. Currently, for simplicity, only some semiempirical models for the transport parameters are integrated into our code as a pre-processed package.
The Thomas-Fermi model is widely used to calculate the ionization degree,
which treats the electrons non-interacting with each other in an atom (or ion) in an effective potential, and works well for metals where the electron density is almost uniform due to strong screening. The Saha-Boltzman model is usually employed in dense plasmas where electrons, ions, and photons are in a local thermal equilibrium state \cite{Hahn97}.
For a gas with single atomic species, the number density $n_i$ of $i$-th ionization state is given by Saha-Boltzman equation
\begin{equation}\label{saha}
\frac{n_en_{i+1}}{n_i}=\frac{2g_{i+1}}{g_i}(\frac{2\pi m_eT_e}{h^2})^{3/2}\exp(-\frac{\phi_{i+1}-\phi_i}{T_e}).
\end{equation}
where $n_e=\Sigma_ii\cdot{n_i}$ is the electron number density, $g_i$ is the degeneracy of state for ions with $i$-th ionization, $\phi_i$ is the ionization potential for ionizing $i$ electrons from a neutral atom in units of eV, $h$ is the Planck constant, $T_e$ is the gas electron temperature.
The ionization potential $\phi_i$ and degeneracy $g_i$ for most of the gases can be found in National Institute of Standards and Technology (NIST) database \cite{NIST}. In dense plasma, the pressure ionization will lower the ionization potential, an effect known as ionization potential depression (IPD). Two distinct theoretical models, the Stewart-Pyatt (IPD-SP) \cite{Stewart66} and Ecker-Kr\"{o}ll \cite{Ecker63} (IPD-EK) models have been developed.
Both models are implemented in the HEETS, and they show similar results for what concerns the ionization degree in the high-Z doped targets. Extending the Saha-Boltzman equation to compounds is straight forward by employing the total electron density for each species.

Once the ionization degree $Z$ is obtained, the resistivity $\eta$ can be calculated by the Lee-More model \cite{Lee84} improved by Desjarlais \cite{Desjarlais01}, which includes the effects of electron-electron, electron-ion, and electron-neutron collisions, as well as the melting process of metals.
The model has been proven to capture the main characteristic of resistivity, and is widely used.

\section{Simulation model}\label{sec:4}
In the following section, we will describe 3D simulations carried out by using HEETS to investigate the influence of high-Z dopant on the fast electron propagation in foam targets.
The simulation box employs $200\times200\times160$ cells with a 1$\mu$m cell size.
Laser parameters closing to the petawatt beam of the Vulcan system \cite{Danson98}, at the Rutherford Appleton Laboratory, are employed in the simulations.
The transverse absorption profile into fast electrons is Gaussian $I(r)=\alpha I_0$exp$(-(r/r_{spot})^2)$, where $\alpha=0.25$ is the laser
absorption efficiency as reported in Ref. \cite{Davies09}, $I_0=3\times10^{19}$ W/cm$^2$ is the laser
peak intensity, $r$ is the radial distance from $x=y=100\mu$m,
and $r_{spot}=10\mu$m is the laser focal spot radius.
The wavelength of the laser is set to $\lambda=1.0 \mu m$.
The duration of the laser pulse is 0.7 ps with a sin$^2$ profile increasing to peak intensity then decreasing with a cos$^2$ profile.
The fast electrons are injected from the left boundary ($z=0$), and are assumed to have an exponential energy distribution $f(E)=\frac{1}{\langle E\rangle}$exp$(-E/\langle E\rangle)$, where $\langle E\rangle$ is the average energy given by the ponderomotive scaling \cite{Wilks92}, i.e, $\langle E\rangle=m_ec^2(\sqrt{1+I(r,t)\lambda^2/1.37\times10^{18}}-1)$, and $\langle E\rangle=1.93$MeV is corresponding to the peak laser intensity.
The initial angular distribution of fast electrons is described as
$f(\theta)=\cos^M(\alpha\theta)$,
where $M=6, \alpha=\frac{1}{\theta_{0}}\cos^{-1}(0.5^{1/M}), \theta_{0}=\pi/6$ is the half width at half maximum (HWHM) of fast electron injection angle as
reported in recent experiments \cite{Green08}.
This indicates that the angular distribution here is not dependent on the mean radial velocity of fast electrons.
It is worth mentioning that the angular distribution could be affected by the radial velocity of fast electron beams for laser interaction with solid targets with large preplasmas \cite{Debayle10}. However, it is found that there is no obvious influence on our conclusion as nonzero radial mean velocity is considered within our parameters.
The target is a polystyrene foam C$_{50}$H$_{50}$
consisting of 50\% C and 50\% H, with a density of 0.4g/cm$^3$ and an initial temperature of 1 eV.
The doped foams are  brominated (Br) polystyrene. The atomic concentration of the Br dopant are 5\% and 10\%, respectively, for C$_{50}$H$_{45}$Br$_{5}$ and C$_{50}$H$_{40}$Br$_{10}$.  In order to keep an identical ion number density with C$_{50}$H$_{50}$, the mass densities of these two doped foams are increased to 0.64g/cm$^3$ and 0.87g/cm$^3$, respectively.
Absorbing boundaries are used for the transverse and longitudinal boundaries both for the particles and electromagnetic fields.

\section{Results and discussions}\label{sec:5}
\begin{figure}\suppressfloats\centering
\includegraphics[width=12cm]{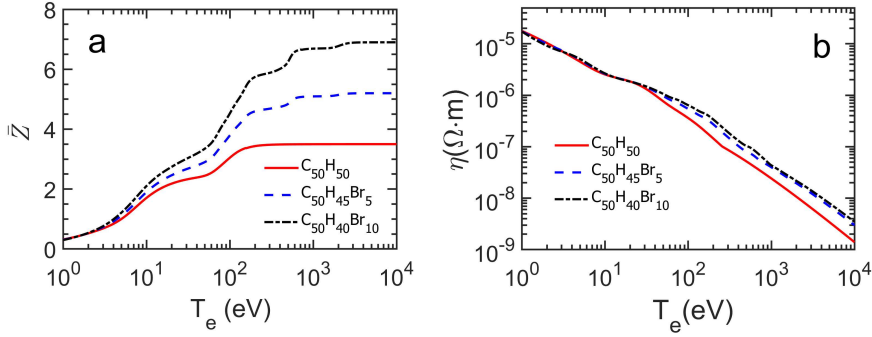}%
\caption{\label{f1} (Color online) Evolution of average ionization degree (a) and resistivity (b) with target temperature for the pure foam, foams doped with $5\%$ and $10\%$ Br, respectively. }
\end{figure}

Figure \ref{f1}(a) shows the evolution of average ionization degree and resistivity with target temperature for the pure foam, foams doped with $5\%$ and $10\%$ Br, respectively.
It is shown that the averaged ionization degree $\bar{Z}$ increases with temperature $T_e$ with a step-like profile due to the distinct ionization energy of each atomic element.
For the pure foam, $\bar{Z}$ shows a rapid increase at low temperature followed by a slower increase, and reaches a constant of $\bar{Z}=3.5$ at $T_e>130$eV due to full-ionization.
Both the saturated ionization degree and the saturation temperature increase with the Br dopant concentration, e.g., $\bar{Z}=5.2$ and $T_e=600$eV for  C$_{50}$H$_{45}$Br$_{5}$, and $\bar{Z}=6.9$ and $T_e=2400$eV for C$_{50}$H$_{40}$Br$_{10}$. Around the lower temperature $T_e=1$eV, the three targets have almost the same ionization degree.

It can be seen that, in Fig. \ref{f1}(b), the resistivity for all the three targets coincide with each other at $T_e<30$eV.
At this relatively lower temperature, the resistivity mainly depends on the electron-neutron collisions as shown in Lee-More model.
For temperature $T_e>100$eV, the Lee-More model reduces to the Spitzer-H\"{a}rm model \cite{Spitzer53}, which considers electron-ion collisions only and scales to the target ionization degree (i.e., $\eta\propto \bar{Z}/T_e^{3/2}$).
Thus, the resistivity of Br doped foam becomes larger than that of the pure foam due to the increase of ionization degree.
As discussed in the introduction, the self-generated magnetic field can grow according to
\cite{Davies97} $\partial_t\vec{B}=\eta\nabla\times\vec{J}_f+\nabla\eta\times\vec{J}_f$, thus the increase of resistivity will lead to an enhancement of magnetic field and collimation of fast electron propagation.

\begin{figure*}\suppressfloats\centering
\includegraphics[width=13cm]{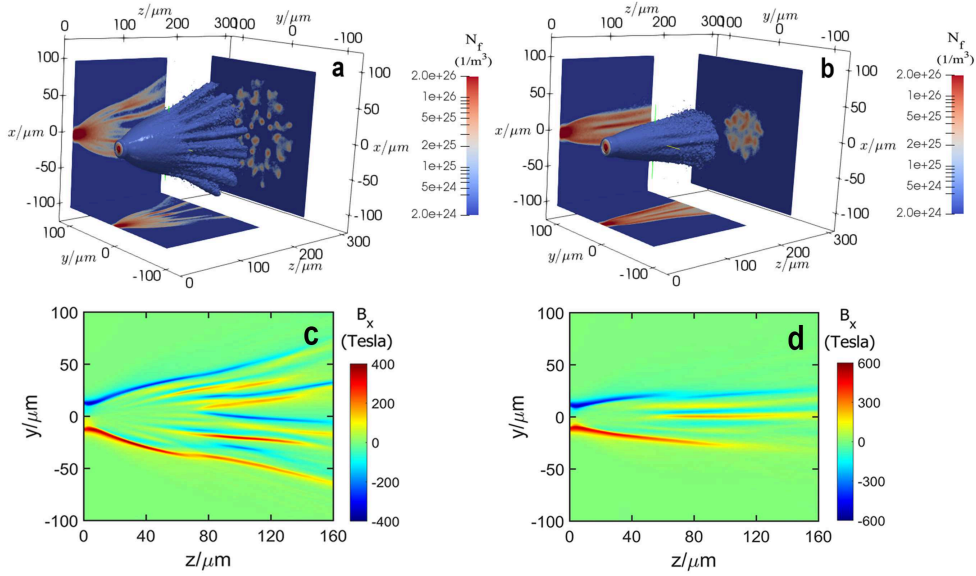}%
\caption{\label{f2} (Color online) Distribution of $log_{10}$ of fast electron density for pure foam (a) and foam doped with $10\%$ Br (b) at t=1.4ps. Corresponding self-generated resistive magnetic field $B_x$ in the $z-y$ plane for pure foam (c) and for foam doped with $10\%$ Br (d).
The electron density and magnetic field are in units of $m^{-3}$ and T, respectively.}
\end{figure*}

The distribution of fast electron density and self-generated resistive magnetic field are shown in Figs. \ref{f2}(a) and (c) at t=1.4ps (0.7ps after the peak of the laser pulse) for fast electron propagation in a pure foam target of density 0.4g/cm$^3$.
Since the resistive magnetic field is weak at early times, which cannot confine the fast electrons, thus the latter propagating in the foam with approximately initial divergence.
As the resistive magnetic fields increase with time, the fast electrons are bent into the high current density region by the resistive field, leading to a reduction in divergence at later times.
A strong filamentation instability driven by the self-generated magnetic field appears deep in the foam.
For clarity, Fig. \ref{f2}(a) also shows the horizontal and vertical slice views of the fast electron density distribution.
The fast electrons can be well confined near the laser injection plane due to the intense magnetic field
at the periphery of the fast electron beam, which reaches $\sim$400T. However, since the magnetic fields get weaker in deeper regions ($z>10\mu m$), the electrons become divergent again.
Distribution of fast electron density at the transverse slice of $z=160\mu m$ is also presented in Fig. \ref{f2}(a).
One can see that more than 30 filaments of fast electrons are induced in the slice. The fast electron current is repelled by the return currents via magnetic force during propagation in targets, inducing the appearance of current filaments, which could become sparser due to the divergent propagation. However, they also could get denser due to the attraction between the filaments. The competition between the attraction and repelling would finally lead to the profiles of the current filaments.

\begin{figure}\suppressfloats\centering
\includegraphics[width=8.5cm]{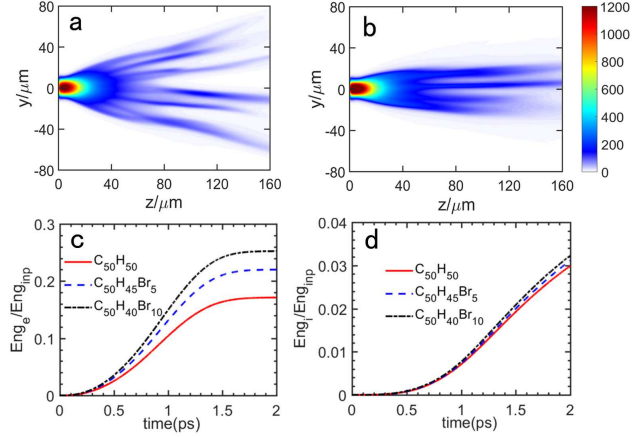}%
\caption{\label{f3} (Color online) Distribution of background electron temperature for pure foam (a) and foam doped with $10\%$ of Br (b) at t=1.4ps. Evolution of the fraction of fast electron energy depositing into background electrons (c) and ions (d) for the pure foam, foams doped with $5\%$ and $10\%$ Br, respectively.}
\end{figure}

In order to confine the fast electrons, we propose here to apply targets containing high-Z dopants, for which the resistivity and density can be modulated and thus the resistive magnetic field can be controlled. To understand the role of high-Z dopant on the fast electron propagation, Figs. \ref{f2}(b) and (d) show the distribution of fast electron density and resistive magnetic field for the C$_{50}$H$_{40}$Br$_{10}$ target.
The fast electrons propagate into the target with their initial divergence in the early stage, similar to what observed in the pure foam. However, they then become more collimated compared to the case of Fig. \ref{f2}(a) due to the stronger resistive magnetic field (which reaches 600T, compared to 400T in the undoped case). The background temperature at the periphery of the fast electron beam is $\sim$100eV for both cases (see Figs. \ref{f3}(a) and (b)), so that the resistivity of C$_{50}$H$_{40}$Br$_{10}$ is about 6 times of that of the pure foam in the range of temperature as shown in Fig. \ref{f1}(b), which induces the stronger resistive magnetic field. It is noted that the number of fast electron filaments is decreased significantly.
At the rear surface of the C$_{50}$H$_{40}$Br$_{10}$ target ($z=160\mu m$), the spot size of the fast electron beam gets much smaller, and the fast electron density becomes much higher than in the pure foam target.

\begin{figure}\suppressfloats\centering
\includegraphics[width=7cm]{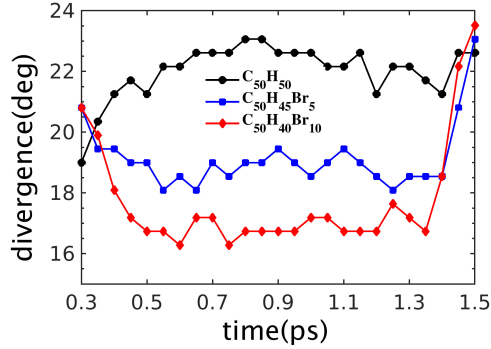}%
\caption{\label{f20} (Color online) Evolution of the mean divergence of fast electrons within the slice of $z=(30\pm1)\mu m$ for the pure foam, foams doped with $5\%$ and $10\%$ Br, respectively. The simulation parameters are same as that in Fig. \ref{f2}}.
\end{figure}

Figure \ref{f20} shows the evolution of the mean divergence of fast electrons around the slice of $z=30\mu m$ for pure foam and doped targets.
The fast electron divergence is estimated by $\theta=\langle\tan^{-1}(p_\perp/p_z)\rangle$, where $p_\perp$ and $p_z$ are the transverse and longitudinal momentum of fast electrons, the average $\langle\rangle$ applies on all the fast electrons arrived at the slice.
The divergence increases rapidly with time for all the three targets due to the fact that fast electrons with small divergence arrive the slice firstly in the early stage ($t<$0.3ps). It increases continuously in the pure foam C$_{50}$H$_{50}$ until $t\sim$0.6ps after which the resistive magnetic can bend the fast electrons.
For the doped targets, the fast electron divergence decreases to a lower value after $t=$0.3ps, e.g., $19^\circ$ for C$_{50}$H$_{45}$Br$_{5}$ and $17^\circ$ for C$_{50}$H$_{40}$Br$_{10}$ around $\sim$0.5ps, as shown in Fig. \ref{f20}.
The discrepancy of the evolution of fast electron divergence in the initial stage in the pure foam and doped targets is attributed to the fact that both the amplitude and spreading width of magnetic field  around the focal spot are increased in the doped targets (not shown for brevity) due to their larger ionization degree and thus the faster growth rate of the magnetic field, which can lead to a better control of the fast electrons (see the follow discussions).
Here the divergence increase after $t=$1.3ps is mainly attributed to the fact that most energetic electrons have penetrated through the slice at this time (recall that the beam duration is 1.4ps) and the remaining electrons have small longitudinal velocity thus relatively large divergence.

In the range of laser and target parameters of our interest, the dominant energy deposition term in Eq. (\ref{tempb}) is Ohmic heating \cite{Davies97,Yang18,Michael95}, and quasi-neutrality is a good approximation, i.e., $\vec{J}_b\simeq -\vec{J}_f$.
Thus the background electron energy equation becomes
\begin{equation}\label{sim0}
C_e\frac{\partial{T_e}}{\partial{t}}= \eta(T_e){J_f}^2,
\end{equation}
where $C_e\simeq1.5n_i\bar{Z}$ is the electron non-degenerate specific heat capacity, $\eta(T_e)= \alpha_0\cdot\bar{Z}{T_e}^{-3/2}$  is the Spitzer-H\"{a}rm resistivity, and $\alpha_0=10^{-4}ln\Lambda$ is nearly constant. Substituting $C_e$ and $\eta$ into the electron energy equation,  one can obtain
\begin{equation}\label{sim1}
\frac{\partial{T_e}}{\partial{t}}=\frac{2\alpha_0}{3n_i}T_e^{-3/2}{J_f}^2.
\end{equation}
This indicates that the electron temperature $T_e$ is only dependent on the fast electron current density and the ion number density, which is not dependent on the ionization degree $\bar{Z}$. It should be noted that Spitzer-H\"{a}rm resistivity is employed to derive Eq. (\ref{sim1}), meaning that it is only valid for targets with high temperature (usually $T_e>$100eV), where the resistivity is mainly determined by the electron-ion collisions.

When resistive diffusion is neglected, the resistive magnetic field $\vec{B}$ is given by
\begin{equation}\label{sim2}
\frac{\partial\vec{B}}{\partial{t}}=-\nabla\times\vec{E}=\nabla\times(\eta\vec{J}_f).
\end{equation}
The azimuthal component of the magnetic field $\vec{B}$ can be estimated as
${\partial{B_\phi}}/{\partial{t}}\simeq\eta J_f/r_{spot}\propto\bar{Z}$ \cite{Yang18,Bell03}.
As the fast electron propagates in the Br-doped foam, the increase of ionization degree $\bar{Z}$ (see Fig.\ref{f1}(a)) will lead to an increase of magnetic field compared to that of the pure foam.

Figures \ref{f3}(a) and (b) show the distribution of background electron temperature for C$_{50}$H$_{50}$ and C$_{50}$H$_{40}$Br$_{10}$ at t=1.4ps, respectively.
The temperature has similar distribution to the fast electron density and resistive magnetic field presented in Fig. \ref{f2}.
According to Eq. (\ref{sim1}), the electron temperature is independent of the ionization degree of the target, i.e., the temperature would be expected to be identical with each other.
However, Fig. \ref{f3}(b) shows that the temperature (with a maximum of 1500eV near the front surface and 300eV around the filaments) in the doped foam is higher than in the pure foam (1200eV and 200eV, respectively, in Fig. \ref{f3}(a)). This discrepancy comes from two factors: 1) both the Spitzer-H\"{a}rm resistivity and the ideal gas specific heat capacity models are good approximations at high temperatures, but will break down at lower temperatures; 2) The enhanced magnetic field can influence the fast electron current density $\vec{J}_f$, so that $\vec{J}_f$ cannot be assumed as a determined constant when solve Eq. (\ref{sim1}).
Actually, since the resistive magnetic field is enhanced in doped foam compared to that in the pure foam, it can lead to a higher current density and thus a higher background temperature.

Figures \ref{f3}(c) and (d) show the fraction of fast electron energy deposited in background electrons and ions for three different targets, i.e., C$_{50}$H$_{50}$, C$_{50}$H$_{45}$Br$_{5}$, and C$_{50}$H$_{40}$Br$_{10}$.
It is seen that energy deposited in the background plasma is increased with increasing dopant concentration of Br, which is mainly due to the increasing Ohmic heating.
The maximum electron density $n_e$ of C$_{50}$H$_{40}$Br$_{10}$ is increased to $\sim2$ times of that of C$_{50}$H$_{50}$ at high temperature ($T_e>$100eV) because of that C$_{50}$H$_{40}$Br$_{10}$ has identical ion number density with C$_{50}$H$_{50}$ but higher ionization.
Though it would lead to an enhancement of collisional heating ($Q_f\propto n_fn_e/T_f^{1/2}$), it is still to be neglected compared to the Ohmic heating (not shown for brevity) and is consistent with that reported in Refs. \cite{Davies97,Yang18,Michael95}.
One can see that, in Fig. \ref{f1}, the target ionization increases with the dopant concentration of Br while the resistivity is almost independent on the Br concentration at lower temperature ($T_e<$30eV). This indicates that the doped target would keep lower temperature for the same injecting fast electron current due to their larger specific heat capacity compared that in the pure foam, and maintain large resistivity for a relatively longer time, which leads to a more efficient Ohmic heating.
Due to the relatively low areal density $\rho R$ of the targets, which is $\rho R=0.04$g/cm$^2$ even for C$_{50}$H$_{40}$Br$_{10}$, the fast electron energy deposited in the target is only 25.7\% and 17\% for C$_{50}$H$_{40}$Br$_{10}$ and C$_{50}$H$_{50}$, respectively.
The background electrons transfer their energy to the ions by collisions after their temperature is increased, and the energy transfer to the ions is negligible($\sim3$ \%), due to the low density of the targets and relatively short time ($2ps$) considered here.

\begin{figure}\suppressfloats\centering
\includegraphics[width=9cm]{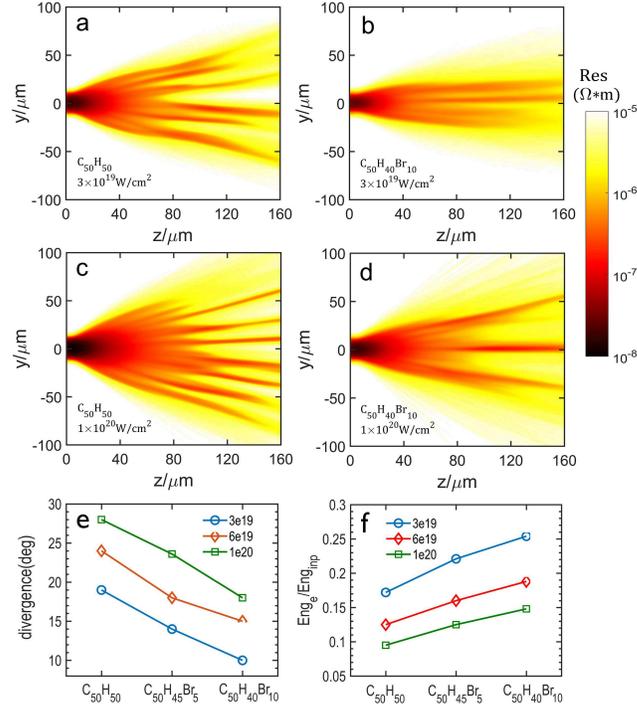}%
\caption{\label{f4} (Color online) Distribution of target resistivity at t=1.4ps, for C$_{50}$H$_{50}$ (a, c) and C$_{50}$H$_{40}$Br$_{10}$ (b, d) of  laser intensities $3\times10^{19}$W/cm$^2$ (a, b) and $1\times10^{20}$W/cm$^2$ (c, d), respectively.
Divergence of fast electrons (e) and the fraction of fast electron energy depositing into background electrons (f) for the cases with different targets and laser intensities. The divergence is extracted from all the fast electrons arrived at the slice of $z=160\mu m$ until t=2ps.}
\end{figure}

For completeness, the effects of high-Z dopant on fast electron propagation at higher laser intensities (I=$6\times10^{19}$W/cm$^2$ and $1\times10^{20}$W/cm$^2$) are also investigated.
Note that though the divergence of fast electrons could increase with laser intensity \cite{Green08}, in order to avoiding influences induced by other parameters, here we are using the same divergence for the fast electrons for different laser intensities. Thus, only the fast electron energy and current density are increased when the laser intensity increases.
The distributions of target resistivity of C$_{50}$H$_{50}$ and C$_{50}$H$_{40}$Br$_{10}$ at $t=1.4ps$ for laser intensity of $3\times10^{19}$W/cm$^2$ are shown in Figs. \ref{f4}(a) and (b), and that for laser intensity of $1\times10^{20}$W/cm$^2$ are shown in Figs. \ref{f4}(c) and (d).
As the fast electrons propagate into the target, the target is heated rapidly, inducing a decrease of the resistivity.
Since larger fast electron current is produced by the higher laser intensity, it heats the background electrons to a higher temperature and leads to a smaller target resistivity compared to that in Figs. \ref{f4}(a) and (b).
According to Eq. (\ref{sim2}), the magnetic field can bend the fast electrons into the region with higher current density $J_f$ (inducing pinching), which also can deflect the fast electrons into the region with higher resistivity $\eta$ (inducing divergence).
Competition between these two effects leads to the propagation profile of the fast electrons in the target.
It is shown that both the transverse width of the resistivity distribution as well as the number of resistive filaments decrease in C$_{50}$H$_{40}$Br$_{10}$, while the width of resistivity distribution spreads widely as the laser intensity increases.

For clarity, Fig. \ref{f4}(e) shows the fast electron divergence for three different targets and three different laser intensities.
The fast electron divergence is extracted from all the fast electrons arrived at the rear surface until the end of simulation time(2ps).
It can be seen that, due to the collimating effect of resistive magnetic field, fast electron divergence is only 19$^\circ$ even in the C$_{50}$H$_{50}$ for the laser intensity of $3\times10^{19}$W/cm$^2$, which is still much smaller than initial divergence (30$^\circ$), and it decreases to 10$^\circ$ in the C$_{50}$H$_{40}$Br$_{10}$ target, indicating that the fast electrons are well collimated.
It is noted that the divergence presented here is smaller than that in Fig. \ref{f20}, which is attributed to the fact that the fast electrons can be bent continuously by the resistive magnetic field after they penetrate through the slice of $z=30\mu m$ where the magnetic field is still sufficiently intense, as shown in Fig. \ref{f2}.
Though the resistive magnetic field is increased for a laser intensity of $10^{20}$W/cm$^2$ because of the higher fast electron current density, it is still not strong enough to collimate the fast electrons due to the average electron energy reaches the value $\langle{E}\rangle=3.88$MeV \cite{Wilks92} (see the discussion in the next paragraph).
The half-divergence angle is about 28$^\circ$ in C$_{50}$H$_{50}$, which approaches the initial angle $30^\circ$.
It will reduce to 23$^\circ$ in C$_{50}$H$_{45}$Br$_{5}$, and further reduce to 18$^\circ$ in C$_{50}$H$_{40}$Br$_{10}$.

The condition for fast electron collimation in the target is derived by Bell \cite{Bell03}, which shows that the ratio of the beam radius $r_{spot}$ to the fast electron Larmor radius $r_L$ should satisfy $r_{spot}/r_L>\theta^2$, so that the fast electron trajectory can be bent by an angle $\theta$ in the distance $r_{spot}/\theta$ and the beam radius approximately doubles.
Combining Ampere¡¯s law, Spitzer resistivity formula, and the background electron temperature equation (considering the Ohmic heating term only), the collimation condition can be described by $\Gamma>1$,
where $\Gamma=0.13n_{23}^{3/5}\bar{Z}^{2/5}\ln\Lambda^{2/5}P_{TW}^{-1/5}T_{511}^{-3/10}(2+T_{511})^{-1/2}r_{\mu m}^{2/5}t_{ps}^{2/5}\theta^{-2}$, $T_{511}$ is the fast electron
temperature in units of 511 keV, $r_{\mu m}$ is the spot radius in units of $\mu m$, $t_{ps}$ is the time in units of picosecond, $P_{TW}$ is the laser power in units of TW, $\theta$ is the fast electron divergence in units of radians, and $\ln\Lambda$ is the Coulomb logarithm.
For a laser intensity of $I=3\times10^{19}W/cm^2$, we set $t_{ps}=1.4$, $\ln\Lambda=5$ for both targets, and $\bar{Z}=3.5$ and 6.9, $n_{23}=1.29$ and 2.54 for C$_{50}$H$_{50}$ and C$_{50}$H$_{40}$Br$_{10}$, respectively, estimated from the target temperature distributions. With these values we obtain $\Gamma=1.1$ for the 10\% Br doped target while $\Gamma=0.56$ for the pure foam target, which is consistent with Bell's collimation condition.

The corresponding fractions of fast electron energy deposited in background electrons are shown in Fig. \ref{f4}(f).
It can be seen that this fraction decreases with an increase of the laser intensity.
According to the conservation of energy flux, i.e. $\alpha I=J_f\langle{E}\rangle/e$, where $e$ is the electron charge,
we can obtain the ratio of fast electron current density for different laser intensities, e.g., $J_{120}/J_{319}\simeq1.667$, where the subscripts `120' and `319' denote laser intensities of $1.0\times10^{20}$W/cm$^2$ and $3\times10^{19}$W/cm$^2$, respectively.
The ratio of the background electron temperature can be estimated as $T_{120}/T_{319}=(J_{120})^{4/5}/(J_{319})^{4/5}=1.505$ using Eq. (\ref{sim1}). Finally, we obtain the ratio of the fractions of electron energy deposition $E_{120}/E_{319}=0.45$, which we can compare to the same quantity calculated from the simulation results: 0.56 (C$_{50}$H$_{50}$), 0.57 (C$_{50}$H$_{45}$Br$_{5}$) and 0.57 (C$_{50}$H$_{40}$Br$_{10}$). The discrepancy is mainly due to the variation of $J_f$ from the pinching or diverging effect of the resistive magnetic field.
In addition, it is found that, for the target doped with 10\% of Br, the fraction of fast electron energy deposited in background electrons only increases slightly, from 17\% to 25.7\% for laser intensity of $3\times10^{19}$W/cm$^2$, and 9\% to 15\% for laser intensity of $1\times10^{20}$W/cm$^2$. The enhanced collimation caused by the doping, coupled to the negligible increase in energy losses to the background, can be very beneficial for applications that require long distance propagation of fast electrons, like fast ignition.

\section{Conclusion}\label{sec:6}
In conclusion, a hybrid PIC/fluid simulation code HEETS has been developed recently. The physical model, numerical methods, as well as transport parameter model are presented in detail.
Laser-driven fast electron beam propagation in foam targets, 5\% and 10\% bromine doped foam targets are then studied by using the hybrid code. It is found that the fast electrons can be better confined in the foam doped with 10\% Br due to its higher ionization and resistivity, which induces a stronger resistive magnetic field to collimate the fast electrons. The increase of energy deposition in targets is small for doped targets, suggesting that this approach could be suitable for propagation of fast electrons over long distances. The scheme is also effective for fast electrons driven by higher laser intensities, as required in fast ignition ($I=10^{20}$W/cm$^2$).
The results here should therefore be helpful for several applications of fast electrons driven by ultra-intense laser pulses.

\ack
This work was supported by the NNSFC (Grant Nos. 11775305, 11305264, 11665012, 11275269, and 11475260) and Science Challenge Project(Grant No. TZ2018001). M.B. acknowledges the support from EPSRC (Grant No. EP/K022415/1). X.H.Y. also acknowledges the support from the China Scholarship Council.

\newpage

\end{document}